\documentclass[aps,11pt]{revtex4}
\usepackage{amsmath,amssymb}
\usepackage{epsfig}
\usepackage{bm}
\usepackage{times}
\usepackage{graphicx}
\usepackage{axodraw}

\begin{document}

\title{NATURAL GAUGE AND GRAVITATIONAL COUPLING UNIFICATION AND 
THE SUPERPARTNER MASSES}

\author{David Emmanuel-Costa}
\email{david.costa@ist.utl.pt}
\author{Pavel Fileviez P\'{e}rez}
\email{fileviez@cftp.ist.utl.pt}
\author{Ricardo Gonz\'{a}lez Felipe}
\email{gonzalez@cftp.ist.utl.pt}

\affiliation{Centro de F\'{\i}sica Te\'{o}rica de Part\'{\i}culas, Departamento de
F\'{\i}sica, Instituto Superior T\'{e}cnico\\  Avenida Rovisco Pais, 1049-001
Lisboa, Portugal}

\date{\today}

\begin{abstract}
The possibility to achieve unification at the string scale in the
context of the simplest supersymmetric grand unified theory is
investigated. We find conservative upper bounds on the superpartner
masses consistent with the unification of gauge and gravitational
couplings, $M_{\tilde G} \lesssim 5$ TeV and $M_{\tilde f} \lesssim
3 \times 10^7$ GeV, for the superparticles with spin one-half and
zero, respectively. These bounds hint towards the possibility 
that this supersymmetric scenario could be tested at future colliders, 
and in particular, at the forthcoming LHC.
\end{abstract}

\maketitle
\pagestyle{plain}
\section{INTRODUCTION}
The unification of all fundamental forces in nature is one 
of the main motivations for the physics beyond the Standard Model. 
More than two decades have passed since the remarkable observation 
that in the minimal supersymmetric extension of the Standard 
Model (MSSM) the gauge couplings unify at a very high-energy
scale~\cite{Dimopoulos:1981yj}, $M_{GUT} \simeq 2 \times
10^{16}$~GeV. Supersymmetric grand unified theories (GUTs) are
considered as the most natural candidates to describe the physics at
the $M_{GUT}$ scale. Nevertheless, this scale is somehow below the
generically predicted perturbative string unification scale $M_{str}
\simeq 5 \times 10^{17}$~GeV~\cite{Ginsparg:1987ee,Kaplunovsky:1987rp}.

Different paths to resolve the discrepancy between the GUT and
string scales have been proposed in the literature~\cite{Dienes:1996du}. 
In particular, the introduction of additional states, with masses 
below the unification scale, is one of the well-motivated 
possibilities. A simple example is provided by the addition of 
adjoint representations such as a color-$SU(3)$ octet ($\Sigma_8$) 
and a weak-$SU(2)$ color-neutral triplet ($\Sigma_3$)~\cite{Bachas:1995yt}. 
In this framework, the role of the adjoint scalars is to push the GUT 
scale up to $M_{str}$. These adjoint scalars are present in 
the $\mathbf{\hat{24}_H}$ representation of the minimal supersymmetric
$SU(5)$~\cite{SUSYSU(5)}, where the MSSM matter superfields are
unified in $\mathbf{\hat{\bar 5}}$ and $\mathbf{\hat{10}}$, and the
Higgs sector is composed of $\mathbf{{\hat 5}_H}$,
$\mathbf{\hat{\bar 5}_H}$ and $\mathbf{\hat{24}_H}$ representations.

As is well known, proton decay is the most dramatic prediction
coming from grand unified theories~\cite{review}. However, it is
interesting to look for alternative ways to test the idea of the
unification of all fundamental forces in nature. In this letter we
investigate if the unification of the gauge and gravitational
couplings at the string scale can give us some new insight in our
quest for unification. We study the possibility to achieve
unification of gauge couplings and gravity in the context of the
simplest supersymmetric grand unified theory. We show that such a
unification leads at one-loop level to a unique relation between the
superpartner masses. Using the electroweak precision data and the
current limits on the SUSY partner masses, we find upper bounds on
the sfermion and fermionic superpartner masses. We conclude that in 
this minimal framework the fermionic superpartner masses are 
naturally at or below the TeV scale.
\section{UPPER BOUND ON THE SUPERPARTNER MASSES}
In this section we will explain the possibility to find upper 
bounds on the superpartner masses once the unification of all 
forces is assumed in the context of heterotic string scenarios.  
In a weakly-coupled heterotic string theory, gauge and gravitational
couplings unify at tree level~\cite{Ginsparg:1987ee},

\begin{equation} \label{unif_rel}
\alpha_{str}=\frac{2\, G_N}{\alpha^\prime}=k_i\,\alpha_i \,,
\end{equation}
where $\alpha_{str}=g_{str}^2/4\pi$ is the string-scale unification
coupling constant, $G_N$ is the Newton constant, $\alpha^\prime$ is
the Regge slope, $\alpha_i=g_i^2/4\pi\, (i=1,2,3)$ are the gauge
couplings and $k_i$ are the so-called affine or Ka\v{c}-Moody levels
at which the group factors $U(1)_Y$, $SU(2)_L$ and $SU(3)_C$ are
realized in the four-dimensional string \cite{Dienes:1996du}.
Including one-loop string effects, the unification scale $M_{str}$
is predicted as~\cite{Kaplunovsky:1987rp}
\begin{equation} \label{Mstring}
  M_{str}=\sqrt{4\pi\alpha_{str}}\:\Lambda_s\,,
\end{equation}
where $\Lambda_s\approx 5.27\times 10^{17}$ GeV.

Our main goal is to investigate the possibility to achieve
unification of all interactions in the context of the minimal
supersymmetric $SU(5)$ theory. The relevant one-loop renormalization
group equations are given by
\begin{equation}
\label{eq:gceq}
  \alpha^{-1}_{iZ}\,=\,k_i\:\alpha^{-1}_{str}\,+\,
  \frac{b_i^{\text{\tiny SM}}}{2\pi}\log\frac{M_{str}}{M_Z} +
  \sum_{R}\frac{\Delta^R_i}{2\pi}\log\frac{M_{str}}{M_R}\,,
\end{equation}
where $\alpha_{iZ} \equiv \alpha_i(M_Z)_{\overline{DR}}$ are the
couplings defined in the $\overline{DR}$ renormalization scheme. The
masses $M_R$ are the different thresholds included in the running.
We recall that in the Standard Model $b_i^{\text{\tiny
SM}}=(41/6,-19/6,-7)$. The coefficients $\Delta^R_i$ are the
additional contributions associated to each mass threshold $M_R$. In
Table~\ref{tab1} we list their values for the minimal SUSY $SU(5)$
theory considered here. In the above equations we have used
$M_{str}$ as the most natural value for the superheavy gauge boson
masses as well as for the mass of the colored triplets in
$\mathbf{{\hat 5}_H}$ and $\mathbf{\hat{\bar 5}_H}$, relevant for
proton decay~\cite{review}. Notice that since the contribution of 
the colored triplets to $b_1 - b_2 \ (b_2 -b_3)$ is positive 
(negative) the upper bounds presented below are the most 
conservative bounds. In other words, the lower the colored 
triplet mass scale is, the lighter the superpartner masses 
have to be to achieve unification.  

The affine levels $k_i$ are those corresponding to the standard
$SU(5)$ theory, i.e. the canonical values $k_1=5/3$, $k_2=1$ and
$k_3=1$. We remark that considering a higher Ka\v{c}-Moody level $k$
(as required, for instance, in string models having a $G\times G$
structure~\cite{stringsGUTs}) simply corresponds to the redefinition
$\Lambda_s \rightarrow \sqrt{k}\,\Lambda_s$. This pushes the string
scale $M_{str}$ up and would require slightly lower values of the
adjoint scalar masses and somewhat heavier sfermions to achieve
unification.

\begin{table}
\begin{tabular}{c@{\ \ }ccc}
\hline
  $R$ & $\Delta^R_1$ & $\Delta^R_2$ & $\Delta^R_3$\\[1mm]
\hline
\\
$\widetilde{G}$ & $2/3$ & $2$ & $2$\\[1mm]
$\widetilde{f}$ & $7/2$ & $13/6$ & $2$\\[1mm]
$\hat{\Sigma}_8 \subset \mathbf{\hat{24}_H}$ & $0$ & $0$ & $3$\\[1mm]
$\hat{\Sigma}_3 \subset \mathbf{\hat{24}_H}$  & $0$ & $2$ & $0$\\[1mm] \hline
\end{tabular}
\caption{The additional contributions to the one-loop beta
coefficients in the context of the minimal SUSY $SU(5)$. Here
$\widetilde{G}$ stands for gauginos and Higgsinos and
$\widetilde{f}$ for sfermions and the extra Higgs doublet.}
\label{tab1}
\end{table}

\begin{figure}[t]
\includegraphics[width=10cm]{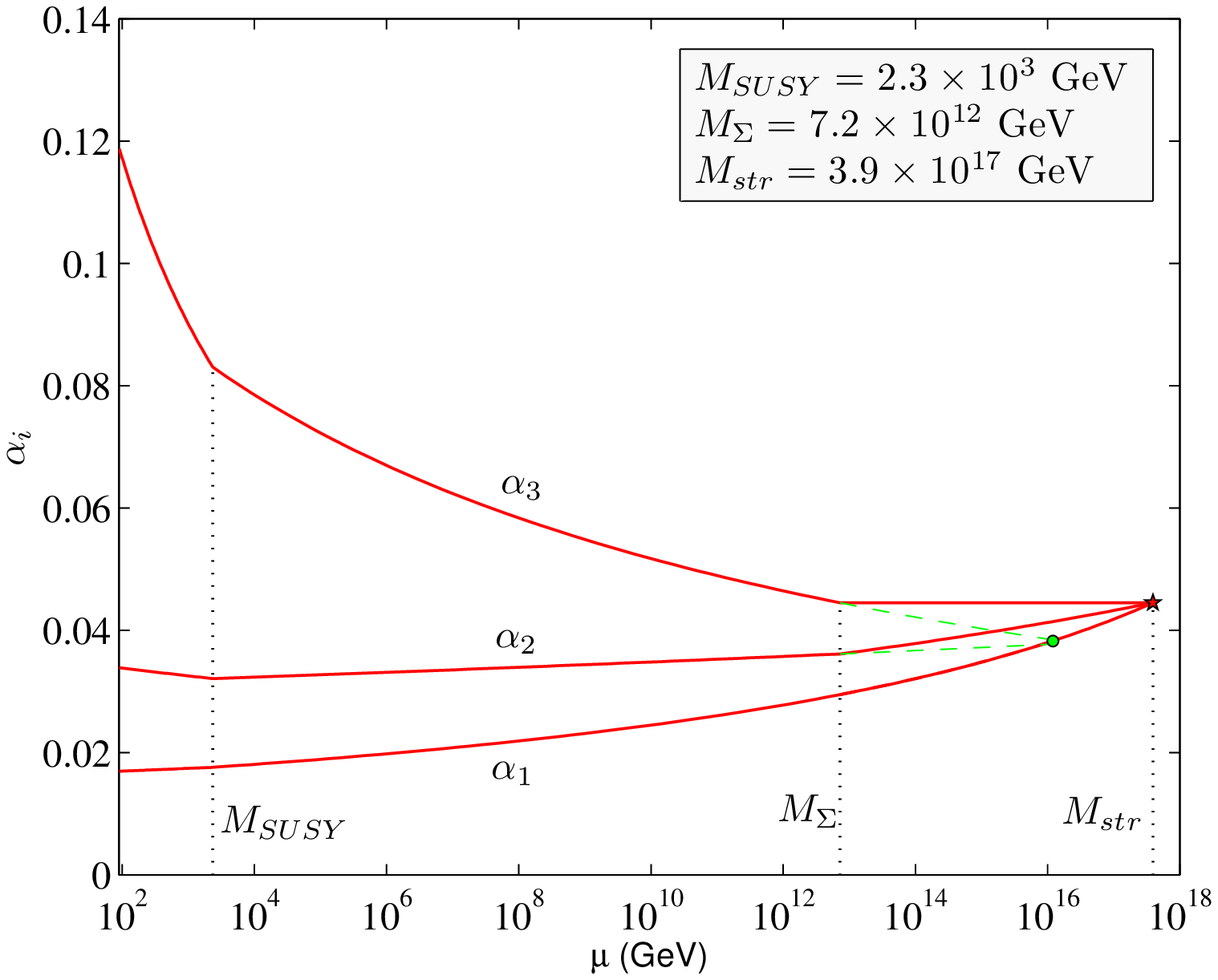}
\caption{\label{fig1} (Color online) The running of the gauge
couplings in the minimal SUSY $SU(5)$ theory for the case of a
degenerate SUSY threshold $M_{\tilde{G}}=M_{\tilde{f}} \equiv
M_{SUSY}$ consistent with the unification with gravity. The dashed
curves correspond to the standard running in the MSSM without 
imposing unification with gravity.}
\includegraphics[width=10cm]{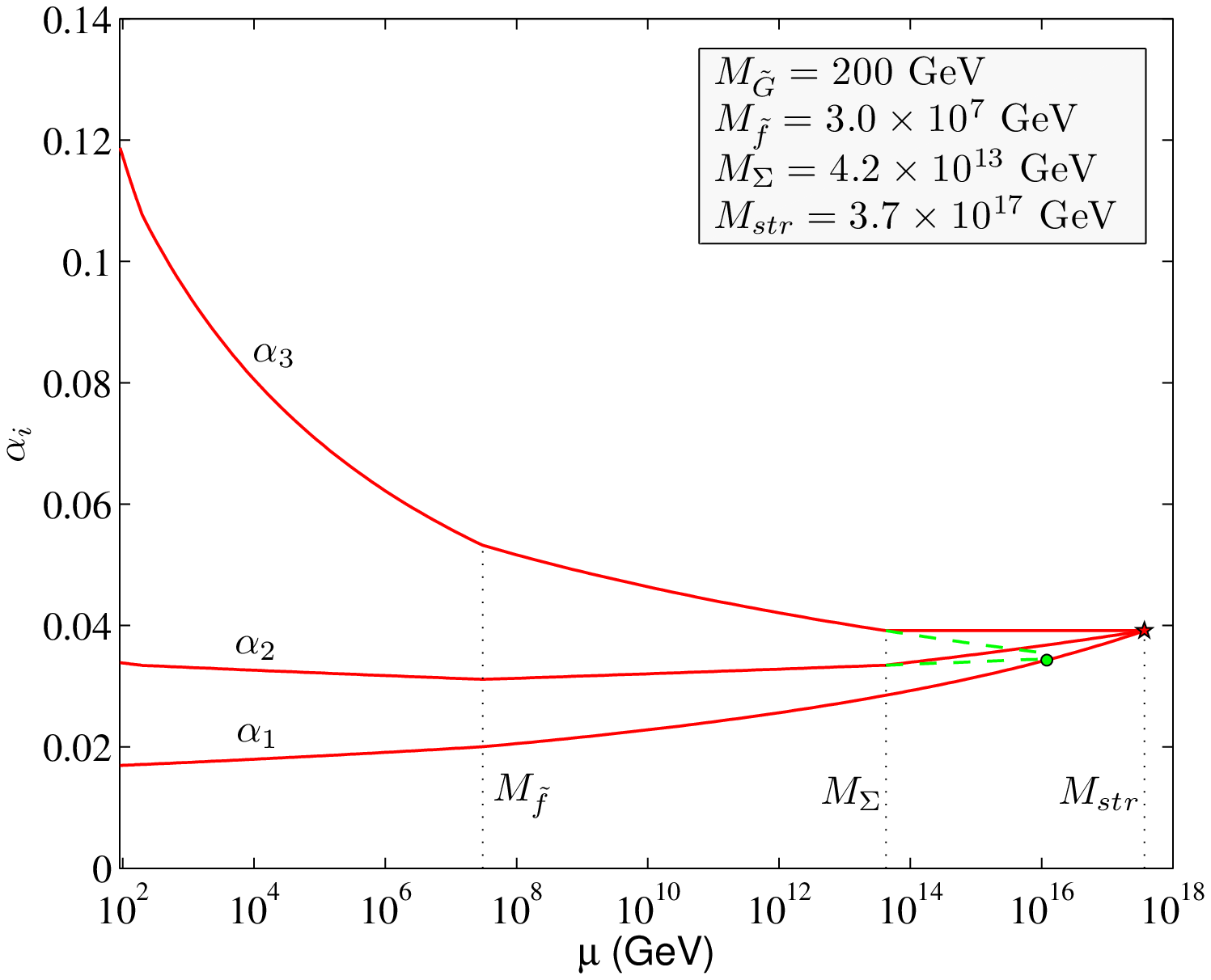} \caption{\label{fig2}
(Color online) Gauge coupling unification in the minimal SUSY
$SU(5)$ theory for a common gaugino mass $M_{\tilde{G}}=200$~GeV
consistent with the unification with gravity. The dashed curves
correspond to the MSSM case.}
\end{figure}

\begin{figure}[h]
\includegraphics[width=10cm]{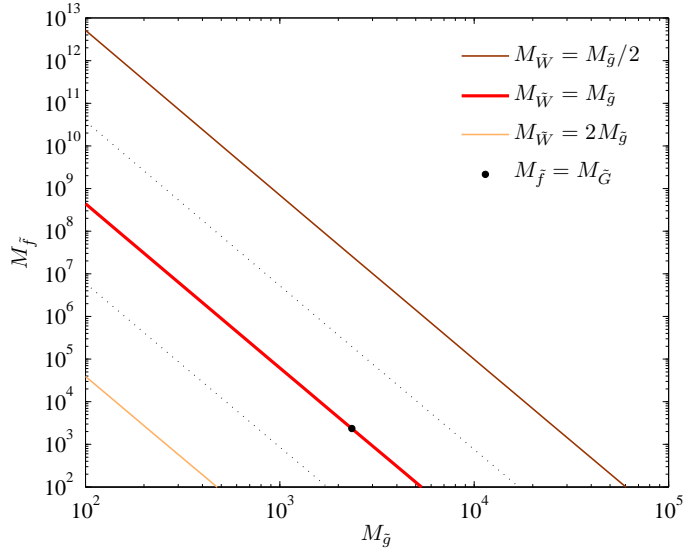}
\caption{\label{fig3} (Color online) Curves (solid lines) in the
($M_{\tilde{g}},M_{\tilde{f}}$)-plane consistent with the
unification of gauge and gravitational couplings for different mass
ratios $M_{\tilde{g}}/M_{\tilde{W}}$. The dotted lines reflect the
$\alpha_s(M_Z)$ experimental uncertainty for the case of degenerate
gaugino masses. The black dot corresponds to a fully degenerate SUSY
partner spectrum at $M_{SUSY}=2.3$~TeV.}
\end{figure}

\begin{figure}[h]
\includegraphics[width=10cm]{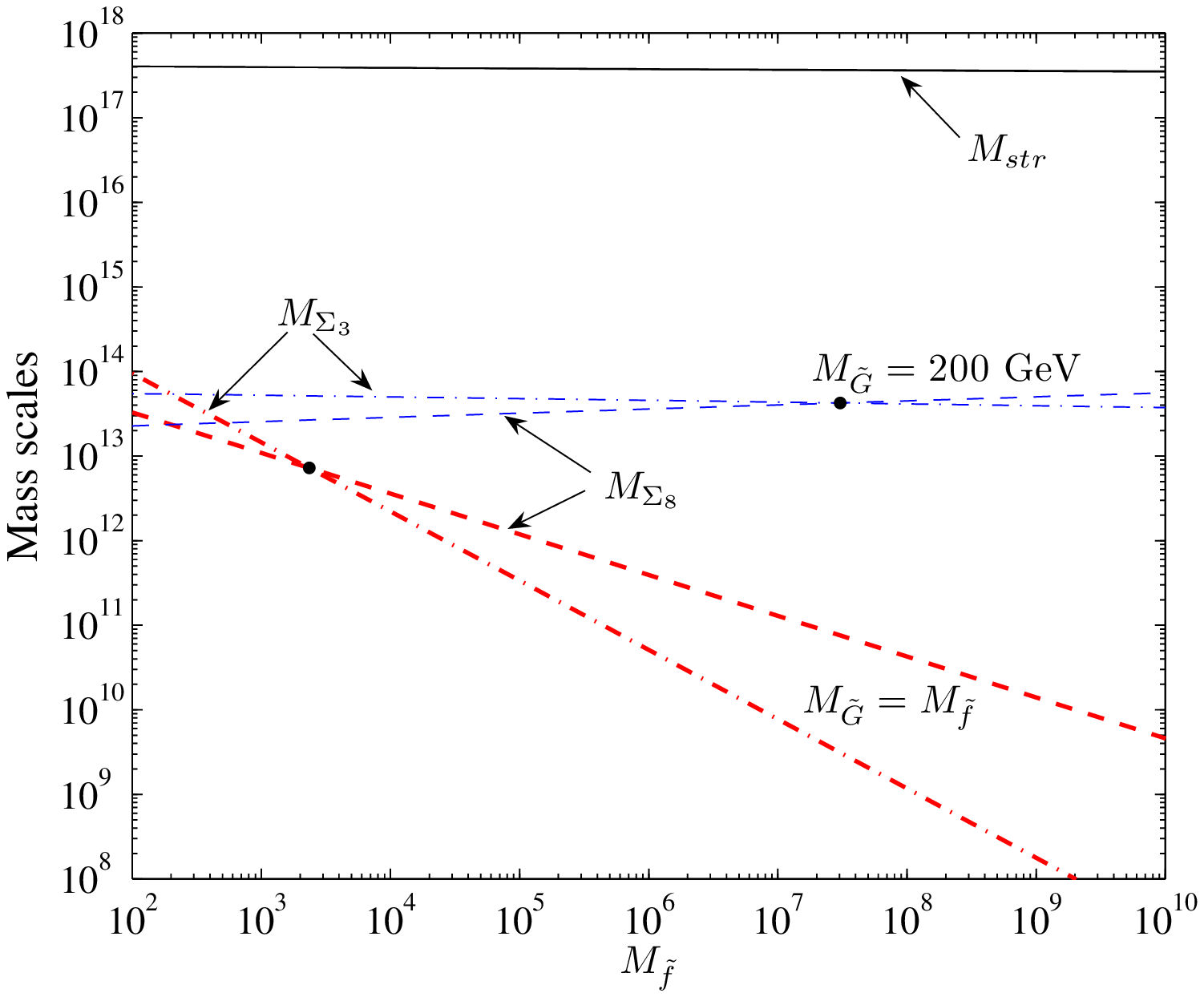}
\caption{(Color online) The dependence of the adjoint scalar masses
$M_{\Sigma_8}$ (dashed lines) and $M_{\Sigma_3}$ (dot-dashed lines)
on the sfermion mass scale $M_{\tilde{f}}$ for two different
scenarios: a low gaugino mass $M_{\tilde{G}}=200$~GeV (blue thin lines)
and a degenerate superpartner mass scale
$M_{\tilde{G}}=M_{\tilde{f}}$ (red thick lines). The black dots correspond
to the solutions presented in Figs.~\ref{fig1} and \ref{fig2} for a
degenerate mass $M_{\Sigma}=M_{\Sigma_3}=M_{\Sigma_8}$.}
\label{fig4}
\end{figure}

Assuming a common mass $M_{\widetilde{G}}$ for gauginos and
Higgsinos, as well as a common mass $M_{\tilde{f}}$ for sfermions
and the extra Higgs doublet, and using $M_{\Sigma_3}=M_{\Sigma_8}
\equiv M_{\Sigma}$ as predicted by the minimal supersymmetric
$SU(5)$ model, the system of Eqs.~(\ref{eq:gceq}) has the solution

\begin{align}
\label{eq:Msfermion} M_{\tilde{f}}& =
\,\frac{M_{str}^6}{M_Z\,M^4_{\tilde{G}}}
\,e^{\pi\left(3\alpha^{-1}_{1Z}
-15\alpha^{-1}_{2Z}+10\alpha^{-1}_{3Z}\right)}\,,\\
M_{\Sigma} & = \,\frac{M_Z^{11/3}}{M_{str}^2\,M^{2/3}_{\tilde{G}}}
\,e^{\pi\left(\frac{1}{2}\alpha^{-1}_{1Z}-
\frac12\alpha^{-1}_{2Z}-\frac13\alpha^{-1}_{3Z}\right)}\,,\label{eq:Msigma}
\end{align}
with the unification scale $M_{str}$ given by
\begin{align} \label{eq:alphastring}
\frac{\Lambda_s^2}{M_{str}^2}= \frac3{8\pi^2}\,W_{0}\left[
\frac{8\pi^2}{3}\frac{\Lambda_s^2 M_Z^{2/3}}{M_{\tilde{G}}^{8/3}}\,
e^{\pi\left(\frac{5}{2}\alpha^{-1}_{1Z}-
\frac{21}{2}\alpha^{-1}_{2Z}+7\alpha^{-1}_{3Z}\right)} \right]\,,
\end{align}
where $W_0(x)$ is the principal branch of the Lambert
function~\cite{Emmanuel-Costa:2005nh,Corless:1996}.

Notice that from Eqs.~(\ref{eq:Msfermion}) and
(\ref{eq:alphastring}) we can find a unique relation between the
gaugino and sfermion masses in this minimal framework. Our main
result reads then as

\begin{align}
\label{main} M_{\tilde{f}}& = \frac{(8\pi^2)^3 \ \Lambda_{s}^6 \
e^{\pi\left(3\alpha^{-1}_{1Z}
-15\alpha^{-1}_{2Z}+10\alpha^{-1}_{3Z}\right)} } {27 M_Z \
M^4_{\tilde{G}} \ W_{0}^3 \left[ \frac{8\pi^2}{3}\frac{\Lambda_s^2
M_Z^{2/3}}{M_{\tilde{G}}^{8/3}}\,
e^{\pi\left(\frac{5}{2}\alpha^{-1}_{1Z}-
\frac{21}{2}\alpha^{-1}_{2Z}+7\alpha^{-1}_{3Z}\right)} \right] }.
\end{align}

In the case when $M_{\tilde{G}}=M_{\tilde{f}} \equiv M_{SUSY}$, from
Eq.~(\ref{eq:Msfermion}) we find that the common superpartner mass
is given by
\begin{equation}
\label{eq:lowsusy} M_{SUSY} =
\,\frac{M_{str}^{6/5}}{M_Z^{1/5}}\,e^{\pi\left(\frac35\alpha^{-1}_{1Z}
-3\alpha^{-1}_{2Z}+2\alpha^{-1}_{3Z}\right)}\,,
\end{equation}
which corresponds precisely to a degenerate SUSY threshold at a
low-energy (TeV) scale. In this scenario, the mass scales $M_{SUSY},
M_{\Sigma}$ and $M_{str}$ are uniquely determined. We find
$M_{SUSY}=2.3$~TeV, $M_{\Sigma}=7.2\times 10^{12}$~GeV and
$M_{str}=3.9\times 10^{17}$~GeV, taking at $M_Z = 91.187$~GeV the
input values $\alpha_s (M_Z)_{\overline{MS}}=0.1176$, $\sin^2
\theta_W (M_Z)_{\overline{MS}}=0.2312$ and
$\alpha^{-1}(M_Z)_{\overline{MS}}= 127.906$~\cite{Yao:2006px}. In
Fig.~\ref{fig1} we show the evolution of the gauge couplings with
the energy scale $\mu$. The role of the adjoint scalars
$\Sigma_{3,8}$ in lifting the unification scale to the string scale
becomes evident from the figure. For comparison, a similar plot is
presented in Fig.~\ref{fig2} for the case of a split-SUSY
scenario~\cite{Split-SUSY} with a common gaugino mass
$M_{\tilde{G}}=200$~GeV, which corresponds to the presently
available experimental lower bound~\cite{Yao:2006px}. In the latter
case, we obtain from
Eqs.~(\ref{eq:Msfermion})-(\ref{eq:alphastring}) the following mass
scales: $M_{\tilde{f}}=3 \times 10^7$~GeV, $M_{\Sigma}=4.2\times
10^{13}$~GeV and $M_{str}=3.7\times 10^{17}$~GeV. Similar results
can be obtained for a higher $k>1$ affine level. For $k=2$ we find
$M_{SUSY}=3.6$~TeV, $M_{\Sigma}=2.7\times 10^{12}$~GeV and
$M_{str}=5.6\times 10^{17}$~GeV for the low-energy supersymmetric
case, while $M_{\tilde{f}}=2.3 \times 10^8$~GeV,
$M_{\Sigma}=2.2\times 10^{13}$~GeV and $M_{str}=5.2\times
10^{17}$~GeV for the split-SUSY scenario with
$M_{\tilde{G}}=200$~GeV.

In our analysis we have assumed a common mass for all superpartners
with the same spin. This is an approximation to a realistic spectrum 
that is produced in several scenarios of supersymmetry breaking as, 
for instance, in models based on minimal supergravity (mSUGRA)~\cite{mSUGRA}. 
Our approximation represents averages of the mass spectra in these models. 
A more realistic analysis of the sparticle masses will not change the 
main conclusions of our work. We may ask ourselves how a mass splitting 
between the superpartners could modify the unification picture. 
In particular, one could expect different masses for 
the gluino ($\tilde{g}$), the weak-gauginos 
($\tilde{W}$) and Higgsinos ($\tilde{h}$). To illustrate the dependence 
of our results on the gaugino spectrum, and without committing 
ourselves to any specific SUSY breaking scenario, we present 
in Fig.~\ref{fig3} the gauge unification curves
(solid lines) in the ($M_{\tilde{g}},M_{\tilde{f}}$)-plane for
different mass ratios $M_{\tilde{g}}/M_{\tilde{W}}$. For simplicity
we have assumed $M_{\tilde{h}}=M_{\tilde{W}}$. From Fig.~\ref{fig3}
we conclude that the present experimental lower bound coming from
sfermion searches, $M_{\tilde{f}} \gtrsim
100$~GeV~\cite{Yao:2006px}, implies an upper bound on the gaugino
masses. Using the central value of $\alpha_s (M_Z)$ we obtain for
$M_{\tilde{G}}=M_{\tilde{g}}=M_{\tilde{W}}$ (solid red line) the
upper limit

\begin{align} \label{boundG}
M_{\tilde{G}} \lesssim 5~\text{TeV}.
\end{align}
Similarly, the experimental lower bound $M_{\tilde{g}}\gtrsim
200$~GeV yield an upper bound on the sfermion scale,
\begin{align} \label{boundf}
M_{\tilde{f}} \lesssim 3\times 10^7~\text{GeV}.
\end{align}

If we take into account the presently allowed $\alpha_s$
uncertainty, then there is a sizable shift of the curves (see dotted
lines). Clearly, these bounds could also be subject to modifications
if gaugino masses are non-degenerate, as can be seen from the
figure. The result of Eq.~(\ref{boundf}) is consistent with the 
upper bound on the scalar masses of $\sqrt{m_{3/2} M_{Pl}}\sim 10^{10}$ GeV~\cite{Kors:2004hz} 
in SUGRA and string models coming from the cancellation of vacuum energy.
We recall that $m_{3/2}$ is the gravitino mass. We also notice that the upper bound 
on the superpartner masses is in agreement with the cosmological 
constraints on the gluino lifetime~\cite{gluino}.

In a similar way, one can consider the case when the adjoint scalars
$\Sigma_3$ and $\Sigma_8$ have different masses~\cite{German}. In
Fig.~\ref{fig4} we present the solutions for $M_{\tilde G}$ and
$M_{\tilde f}$ consistent with unification. We notice that when the
mass splitting is small the fermionic superpartner masses in agreement with
unification are in the interesting region for LHC. However, if we
restrict ourselves to the minimal supersymmetric $SU(5)$, where
these adjoint fields have to be degenerate, the upper bounds given
in Eqs.~(\ref{boundG}) and (\ref{boundf}) hold. 

Let us also comment on some other relevant effects. As explained before, 
when the colored triplets in $\mathbf{{\hat 5}_H}$ and 
$\mathbf{\hat{\bar 5}_H}$ are below the unification scale the 
masses of the superpartners have to be smaller. Therefore, the upper 
bounds on the superpartner masses are indeed those coming from 
the case when the colored triplets are at the unification scale. 
String threshold effects as well as two loop effects have been 
neglected in our analysis. These effects could be important and 
we will be studied elsewhere. However, as we have pointed out, there 
are other relevant effects at one-loop level, such as the 
mass splitting between the fermionic superpartners, which already 
indicate that only in the simplest scenario conservative upper 
bounds on the superpartner masses can be found.       

\section{Summary}
We have investigated the possibility to achieve unification of the 
gauge and gravitational couplings at the perturbative string scale 
in the context of the simplest supersymmetric grand unified theory. 
We have pointed out a unique one-loop relation between the superpartner 
masses consistent with the unification of all interactions. 
Conservative upper bounds on the superpartner masses were found, 
namely, $M_{\tilde G} \lesssim 5$ TeV and $M_{\tilde f} \lesssim 3 \times 10^7$ GeV, for the spin-1/2 and spin-0 superpartners, respectively. 
These bounds hint towards the possibility that this supersymmetric 
scenario could be tested at future colliders, and in particular, 
at the forthcoming LHC.

\begin{acknowledgments}
{\small We would like to thank Pran Nath for the careful 
reading of the manuscript 
and very useful comments. The work of D.E.C. and R.G.F. has
been supported by {\em Funda\c{c}\~{a}o para a Ci\^{e}ncia e a Tecnologia} (FCT,
Portugal) under the grants \mbox{SFRH/BPD/1598/2000} and
\mbox{SFRH/BPD/1549/2000}, respectively. P.F.P. is supported by FCT
through the project \mbox{CFTP-POCTI-SFA-2-777} and a fellowship
under the project \mbox{POCTI/FNU/44409/2002}. P. F. P would like to
thank I. Dorsner, C. Nu\~{n}ez, A. Uranga for comments and G. Walsch for support.}
\end{acknowledgments}


\end{document}